\documentclass[11pt]{article}
\pdfoutput=1
\usepackage{graphicx,subfigure,bm,color,hyperref}
\usepackage{amsfonts}
\usepackage{color}

\begin{document}
	
	\begin{center}
		\large{\bf{Duality Between Dirac Fermions in Curved Spacetime \\
		and Optical solitons in Non-Linear Schrodinger Model:\\Magic of $1+1$-Dimensional Bosonization  }}	
	\end{center}
	\vspace{0.0cm}
	\begin{center}
		
		Subir Ghosh \footnote{E-mail: subirghosh20@gmail.com }\\
		\vspace{2.0 mm}
		\small{\emph{Physics and Applied Mathematics Unit, Indian Statistical
				Institute\\
				203 B. T. Road, Kolkata 700108, India}} \\
	\end{center}
	\vspace{0.5cm}

\begin{abstract}
Bosonization in curved spacetime maps massive Thirring model (self-interacting Dirac fermions) to a generalized sine-Gordon model, both living in $1+1$-dimensional  curved spacetime. 	Applying this duality we have shown that the Thirring model fermion, in non-relativistic limit, gets identified with the soliton of  non-linear 	Scrodinger model with Kerr form of non-linearity. We discuss one particular optical soliton in the latter model and relate it with the Thirring model fermion.
\end{abstract}	

\vspace{1.0cm}	  
	
{\bf Introduction :} Mapping between distinct  	theories has proved to be a powerful tool throughout the development of physics. Apart from the aesthetic satisfaction of relating very different physical systems via mapping their respective theories, it has  great analytical and experimental utility. In some cases, the duality is between a strongly interacting theory and a weakly interacting theory so that analytical results from the latter (that are easier to derive) can yield important results for the former (that may be intractable as such). In this context, the most celebrated example of recent times is the $AdS-CFT$ correspondence \cite{mal} where the duality is between a weakly coupled theory (gravitation in the bulk) and a strongly coupled theory (conformal field theory at the boundary). Another form of useful duality exists whereby two different systems are governed by  formally similar dynamics, (eg. structurally same  equations of motion in a purely algebraic sense) such that one of the systems is more conducive to experiments that can provide information of verification of certain conjecture in the dual theory. A very well-known  example in this context is the analogue gravity models \cite{review} that provide signatures of Hawking radiation which is operationally impossible to detect in  realistic General Relativity scenario, mainly due to its exceedingly small value. 

In a recent paper \cite{jules} a novel form of duality has been revealed between a system obeying Dirac equation in $(1+1)$-dimensioal curved spacetime and the multiphoton Rabi model. Exploiting this duality it is possible to simulate behavior of a relativistic particle near a strong gravitational field (Dirac equation for the particle in a black hole background) in a trapped-ion experimental platform (Rabi model). Interestingly enough, experimentally observed ( as well as analytically computed in  Rabi model framework) nature of the particle trajectory matches with the numerically computed Zitterbewegung motion of the particle in gravitational background. This duality is a result of the algebraic similarity between the two quantum mechanical models where it was sufficient to compare their Hamiltonians. In the present article we further pursue this idea and reveal that there exists another duality at a different level - the quantum field theoretic framework. In particular we have shown that a relativistic theory of self-interacting massive Dirac fermions (massive Thirring model) in curved spacetime can be mapped to a form of non-linear Scroedinger equation supporting Kerr-type of solitons, in the non-relativistic limit. The scheme that allows an exact quantum correspondence between the fermionic theory (massive Thirring model) and the relativistic bosonic theory (Sine-Gordon theory from which the Schroedinger theory emerges in the non-relativistic limit) in $1+1$-dimensions is known as bosonization \cite{col,mand}.

Before proceeding further it is necessary to put the present work in its proper perspective. Our aim is to study fermionic field theory and in particular behavior of fermions in a strong gravitational field. As we have demonstrated in the present work there is a duality between the above theory and Kerr solitons. Thus we can exploit the well studied theoretical literature on Kerr soliton. More importantly because Kerr solitons have been experimentally observed in laboratory one can obtain predictions for the dual fermion theory.  For this reason at the end of this work we have shown how one can map the results from soliton sector to results (or predictions) in the fermion sector. This scheme is very much in line with the work in \cite{jules}.

Generically solitons are finite energy and localized solutions of non-linear differential field equations of motion with an important property that their profiles remain unchanged even after collisions during their time evolution. Kerr solitons are produced as a result of a delicate balance between non-linear Kerr effect (that changes refractive index of an optical medium under applied electric field) and dispersion of the wave propagation.  We mention just a few works on theoretical aspects of Kerr solitons \cite{opsol1} :  Kerr Soliton collisions have been studied in  \cite{zak,seg}  and references therein. A  new kind of  optical  soliton, named as  holographic soliton was proposed in \cite{cohen}. This type of  soliton is formed when two interacting waves generate a  periodic change in the refractive index, and at the same time the solutions are Bragg diffracted  from the same induced grating. 

 The  first experimental finding  Kerr soliton was reported in  \cite{bart}. Other experimental
studies on  Kerr
solitons include collisions \cite{ait,shi,tik} , fission and annihilation \cite{krol},
and spiraling \cite{shi1}. optical solitons have been observed in different systems such as  atom vapor \cite{bj}, photorefractive \cite{du} and photovoltaic crystals \cite{ta},  thermal nonlinearities \cite{sw}, liquid crystals \cite{pe}.

 {\bf Bosonization :} The phenomenon of  bosonization  \cite{col,mand}, an exact duality between a relativistic fermionic theory and a relativistic bosonic theory, can be achieved in $1+1$-dimension. In the present instance, the Dirac theory in curved background can be mapped to a generalized sine-Gordon theory in curved spacetime \cite{eboli}. In this paper we have shown that the latter, in non-relativistic limit and for small coupling reduces to Non-Linear Schrodinger (NLS) model with Kerr form of non-linearity \cite{opsol}, allowing various types of optical soliton solutions (for a review, see \cite{opsol1}). We are interested in this final identification and will reveal the connection between the original Dirac fermion in gravity and the NLS optical soliton.

  In the original bosonization, Coleman \cite{col} considered  a self-interacting fermion theory, the massive Thirring model in flat spacetime
  \begin{equation}
  L_{MT}=\frac{1}{c}[i\hbar c\bar \chi \gamma_\mu \partial^\mu \chi -\frac{\lambda^2}{2}j_\mu j^\mu - c^2 m\bar\chi \chi ],~~j_\mu = \bar \chi \gamma_\mu  \chi .
  \label{1}
  \end{equation}
  The fields and parameters have dimensions \\$[\chi^2]=(length )^{-1},~[\lambda^2]=[(mass)(length)^3/(time)^2] $.
  On the other hand the self-interacting bosonic sine-Gordon 	model is
  \begin{equation}
  L_{SG}=\frac{1}{c}[\frac{1}{2}\partial_\mu \phi \partial^\mu \phi +\frac{\alpha}{\beta^2}cos(\beta\phi )],
  \label{2}
  \end{equation}
  with the fields and parameters having dimensions $[\phi^2]=(mass)(length)^3/(time)^2],~[\beta ]=[\phi]^{-1},~[\alpha ]
=(length)^{-2}$.  By comparing the two respective perturbation series term by term, Coleman \cite{col} proved that the fermionic and bosonic theories were equivalent (or dual) provided the following mapping between the coupling constants and the fields are assumed: 
  \begin{equation}
  \frac{4\pi}{c\hbar \beta^2}=1+\frac{\lambda^2}{c\hbar \pi},~~ -\frac{\beta}{2\pi}\epsilon^{\mu\nu}\partial_\nu \phi =j^\mu,~~\frac{\alpha}{\beta^2}cos(\beta\phi )=-\sigma c^2 m\bar\chi \chi
  \label{3}
  \end{equation}
  with $\epsilon ^{01}=1$. $\sigma$ is a cut-off dependent dimensionless numerical parameter \cite{col}.
 Subsequently, Mandelstam \cite{mand} complimented this by constructing Fermi field operators as  non-local combination of canonical Bose fields. The final outcome of this beautiful duality is that the massive Thirring fermion is the sine-Gordon  soliton \cite{col}. The bosonization technique has been generalized for curved spacetime by Eboli \cite{eboli} with a non-trivial result that the sine-Gordon model has a {\it position dependent} effective mass.
 
 {\bf Bosonization in curved spacetime :} Let us provide the results for bosonization in curved spacetime as derived by Eboli \cite{eboli}. The Curved spacetme Massive Thirring model is 
 \begin{equation}
  L_{CMT}=\frac{\sqrt{-g}}{c}[i\hbar c e^\mu _a(\bar \chi \gamma^a \partial_\mu \chi ) -\frac{\lambda^2}{2}j_\mu j^\mu - c^2 m\bar\chi \chi ],~~j_\mu = \bar \chi \gamma_\mu  \chi ,
 \label{4}
 \end{equation}
 where the zweibin fields $(e^\mu _a, e_\mu ^a)$ are defined by $g_{\mu\nu}=e_\mu ^a e_\nu ^b \eta_{ab}$. We exploit the property that any two-dimensional metric can be represented as 
 \begin{equation}
 g_{\mu\nu}(x)=\Omega^2(x)\eta_{\mu\nu}
 \label{5}
 \end{equation}
 with the zweibein fields expressed as $e_\mu ^a=\Omega \delta^a_\mu ,~e^\mu _a=\Omega^{-1} \delta ^\mu _a$ and $\Omega $  being the conformal factor. Following \cite{eboli} we write down the action for the generalized   Sine-Gordon model after bosonization:
\begin{equation}
A_{CSG}= \int dx~dt \sqrt{-g}[ \frac{1}{2}g^{\mu\nu}\partial_\mu \phi\partial_\nu \phi + \frac{\alpha}{\beta ^2 \Omega (x)} cos(\beta \phi  )]
\label{6}
\end{equation}	
or its equivalent form, (to be used subsequently in this work),
\begin{equation}
	A_{CSG}=\int dx~dt [\frac{1}{2}\eta^{ab}\partial_a \phi\partial_b \phi + (\frac{\alpha}{\beta^2} \frac{\sqrt{-g(x)}}{\Omega (x)}) cos(\beta \phi)] .
\label{7}
\end{equation}	
In the present instance the above action turns out to be 
\begin{equation}
	A_{CSG}=\int dx~dt [\frac{1}{2}\eta^{ab}\partial_a \phi\partial_b \phi + (\frac{\alpha}{\beta^2} \Omega (x)) cos(\beta \phi)] .
	\label{7a}
\end{equation}
We use the Minkowski signature $\eta^{00}=1=-\eta^{11}$. This is the generalized sine-Gordon model with position dependent coupling. $M(x)$,  the effective mass of $\phi$, is   given by $\alpha\Omega (x) = (\frac{M(x)c}{\hbar})^2$). (For a physical understanding of this generalization see \cite{eboli}.) The soliton solutions of the conventional (constant coupling) sine-Gordon model are well studied and it should be possible to extend them for the position dependent interacting case, at least in a perturbative framework. Finally note that in  flat spacetime the relation $\alpha/\beta^2 =m' \propto m$, the Thirring fermion mass,  holds. The equation of motion follows:
\begin{equation}
\eta^{ab}\partial_a \partial_b \phi + (\frac{\alpha}{\beta} \Omega (x)) sin(\beta \phi) =0.
\label{7a}
\end{equation}
{\bf Non-relativistic limit and Schrodinger equation :}

Let us consider a non-relativistic reduction of (\ref{7}) via the substitution,
\begin{equation}
	\phi = \frac{1}{\sqrt{2cE_0}}(e^{-iE_0t/\hbar}\psi+e^{iE_0t/\hbar}\psi^*) ,
	\label{9}
\end{equation}	
where $E_0=Mc^2$ is the rest energy.   Upon substitution of (\ref{9}) the terms in (\ref{7}) reduce to,
\begin{equation}
\phi^*\phi =\phi^2\approx \frac{\psi^*\psi}{cE_0},~~ (\partial_x\phi )^2 \approx \frac{\partial_x\psi^*\partial_x\psi}{cE_0},~~ 
\dot\phi ^2 \approx \frac{1}{cE_0}[\frac{E_0^2}{\hbar^2}\psi^*\psi + i\frac{E_0}{\hbar}(\dot \psi\psi^*-\psi\dot{\psi}^*)],
\label{7c}
\end{equation}
where, as it is customary in non-relativistic limit (see eg. \cite{guth}), we have dropped terms quadratic in time-derivatives and terms having time variation $\sim exp(\pm 2iE_0t/\hbar )$.

Since this approximation constitutes an important aspect of our analysis let us elaborate on it a little to understand its physical implication. One may expect that in the nonrelativistic limit, 
fluctuations of a field, oscillating rapidly on time scales $\Delta t$ large compared to inverse rest energy (or effective mass $M$ of  the
scalar field) average to zero, i.e. when $\Delta t >> M^{-1}$. On the other hand, a careful and systematic analysis by \cite{guth,gu1} (see also \cite{eff}) shows that due to  nonlinear self-couplings, the rapidly oscillating terms can act as effective  backreaction  on the dominant,
slowly varying component of the nonrelativistic field. Looking at it from path integral perspective, removal of  the fast oscillating terms  is equivalent to integrating out the high-frequency components of a field. We emphasize that for our purpose the above arguments are adequate since we are interested in deriving non-relativistic low energy soliton solutions and neglect back reaction effects. Thus we recover the correct non-relativistic action,
\begin{equation}
A_{NRCSG}=\int dx~dt \big[\frac{1}{2cE_0}\{\frac{1}{c^2}(\frac{E_0^2}{\hbar^2}\psi^*\psi + i\frac{E_0}{\hbar}(\dot \psi\psi^*-\psi\dot{\psi}^*))-(\partial_x\psi^*\partial_x\psi)\}$$$$+\frac{\alpha\Omega}{\beta^2}cos(\beta\sqrt{\frac{\psi^*\psi}{cE_0}})\big].
\label{7d}
\end{equation}
It needs to be stressed that the origin of the ambiguity in developing non-relativistic limit for a real scalar field theory and its resolution, that is to derive the above action  in a rigorous way is elaborated in \cite{rb}.

 A straightforward variation of $\psi^*$ generates the (Schrodinger) equation of motion for $\psi$, 
\begin{equation}
i\hbar\dot\psi=-\frac{\hbar^2}{2(E_0/c^2)}\partial^2_{x}\psi 
-\frac{E_0}{2}\psi+\frac{\alpha\hbar^2 c^3}{2\beta}\frac{\Omega}{\sqrt{cE_0\psi^*\psi}}
	sin(\beta\sqrt{\frac{\psi^*\psi}{cE_0}})\psi $$$$
	= -\frac{\hbar ^2}{2M}\partial_x^2\psi - \frac{Mc^2}{2}\{1 -\beta\Omega \sqrt{\frac{Mc}{\psi^*\psi}} sin(\beta\sqrt{\frac{\psi^*\psi}{Mc}})\}\psi =0.
\label{7e}
\end{equation}
 This is a very complicated form of non-linear Schrodinger equation. We will study a truncated version of it following an expansion in powers of $\beta$ up to $O(\beta^2)$. Expansion of the $sin$-function yields
  $$ sin(\beta\sqrt{\frac{\psi^*\psi}{Mc}})\approx (\beta\sqrt{\frac{\psi^*\psi}{Mc}})-\frac{1}{6}(\beta\sqrt{\frac{\psi^*\psi}{Mc}})^3.$$
 This considerably simplifies the  Schrodinger equation (up to $O(\beta^2)$) leading to
\begin{equation}
i\hbar \dot \psi =  -\frac{\hbar ^2}{2M}\partial_x^2\psi -\frac{\beta^2}{12c}(\psi^*\psi)\psi.
\label{13}
\end{equation}
Note that, similarly to the  free massive Klein-Gordon field $\phi$ in flat spacetime  in the non-relativistic limit, the term linear in $\psi$ in (\ref{13}) cancels out  in curved spacetime as well.

One point might require some clarification. It appears that a similar looking equation of motion as (\ref{13}) could have been obtained directly from the original fermionic Thirring model (\ref{4}). But one has to keep in mind that classical limit of the Thirring fermion field are anticommuting and their coherent collection as soliton is not possible due to Pauli exclusion principle (unlike the boson field where coherent collection of boson particles behave as classical wave). In fact  the sine-Gordon soliton, (or its non-relativistic reduction to  Kerr soliton), obtained after bosonozation has no direct connection with the basic fermionic Thirring field since the Mandelstam construction \cite{mand} explicitly shows that fermionic field is a non-local and non-perturbative construction of the bosonic soliton field whereas Coleman had shown \cite{col} that the soliton field is mapped to a composite fermionic field (see eg. \cite{orf}) . The map is given in (\ref{3}).

(\ref{13}) is a widely studied form of Non-Linear Schrodinger equation (NLSE), known as Kerr law non-linearity, that appears in water waves and fibre optics \cite{opsol} (for a review see \cite{opsol1}). The distinct types of soliton solutions are known as dark, dark-singular, and bright, bright-singular and also dark-bright and dark-bright singular solitons \cite{opsol}. Hence the curved spacetime bosonozation along with a  non-relativistic reduction has revealed a duality between massive Thirring model fermion in curved spacetime and  optical soliton  of Kerr NLSE. In the present work we will explicitly consider the bright optical soliton as an example.

A generic soliton solution is given by 
\begin{equation}
\psi (x,t)=u(\xi) e^{i\Phi(x,t)};~~\xi =\lambda (x+vt),~~\Phi = -kx+\omega t
\label{a1}
\end{equation}
where $v$ and $\lambda$ respectively  represent velocity and inverse width of the soliton with $k,\omega$ being the wave number and frequency.  Explicit  solution for the bright optical  soliton is given by
\begin{equation}
\psi (x,t)=\frac{B}{\sqrt \hbar} sech[\pm \frac{B}{\sqrt c}\sqrt{\frac{\nu}{2\mu}}(x+2\mu kt)]\times e^{i\{-kx+\frac{1}{2}(-2\mu k^2 +\nu \frac{B^2}{c} )t+\theta \}}
\label{a2}
\end{equation}
where the parameters are
\begin{equation}
\mu = \frac{\hbar}{2M},~\nu=\frac{\beta}{12c\hbar }
\label{a3}
\end{equation}
and the inverse soliton (effective) width $\Lambda$ and dispersion relation are
\begin{equation}
\Lambda = B\sqrt{\frac{\nu}{2c\mu}},~\omega = \frac{1}{2}(-2\mu k^2+\frac{B^2\nu}{c} ).
\label{a4}
\end{equation}
$B,\theta$ are fixed by initial conditions.  It can be checked that $\Lambda$ and $\omega$ have dimensions of $length^{-1}$ and $time^{-1}$ respectively.

{\bf Semi-classical gravity :} So far we have not specified the metric of our curved spacetime. It is well-known that all two-dimensional metrics are conformaly flat which reduces the Einstein tensor to vanish identically. However, there are interesting modifications of Einstein gravity and the one we will consider here is {\it semiclassical gravity} in $1+1$-dimensions for a static point source \cite{mann}. Originally formulated by Jackiw and by Teitelboim \cite{jt}, this model has created a lot of interest very recently due to its relevance in the context of quantum complexity conjecture \cite{su} and its connection to Sachdev-Ye-Kitaev model \cite{syk}. In two dimensions  Einstein gravity action is topological in nature and does not generate a dynamical theory as the Einstein tensor vanished identically. Thus the simplest non-trivial gravity-like theory in  two dimensions is provided by the Jackiw-Teitelboim formulation of gravity that exhibit black hole solutions. In this model, in a modified Einstein equation the Ricci scalar is directly equated with the trace of energy-momentum tensor (and cosmological constant).

The metric is given by 
\begin{equation}
g_{\mu\nu}(x)=diag\{f(x),-f^{-1}(x)\},~~f(x)=\frac{2GM_0}{c^2} | x | +\epsilon =k_{eff}|x|+\epsilon
\label{13a}
\end{equation}
where $k_{eff}=2GM_0/c^2$, $M_0$ denotes the mass of the point source and $\epsilon =  1$ or $\epsilon =  -1$ refers to a naked singularity or a black hole horizon respectively. From $1+1$-dimensional Newton's law the dimension of $G$ is $[G]=length/(mass ~time^2)$ \cite{jules,mann}.  The conformal factor $\Omega = f$ and for singularity with horizon, {\it i.e.} black hole,  we have assumed $x$ to be positive since in one spatial dimension particles can not cross the singularity. Furthermore, we also assumed that $x$ is away from the horizon.
   
  {\bf Connecting with original  fermion (Thirring) model :} Armed with the bosonization dictionary (\ref{3}), the relation (\ref{9}) and finally the solution (\ref{a2} - \ref{a4}), we can retrace our path to get an approximate behavior of the fermion conserved current $j_\mu$ as given in (\ref{1}). $\phi$ is given by
   \begin{equation}
\phi (x,t)=\frac{2B}{\sqrt{2cE_0}} sech[ B\sqrt{\frac{\nu}{2c\mu}}(x+2\mu kt)]$$$$\times cos{\{-kx+\frac{1}{2}(-2\mu k^2 +\nu \frac{B^2}{c} -2\frac{E_0}{\hbar})t+\theta \}}.
\label{p1}
\end{equation}
Exploiting (\ref{3}) we recover the fermion density and current profiles,
\begin{equation}
j^0=-\frac{\beta}{2\pi}\partial_x \phi ;~~j^x=\frac{\beta}{2\pi}\dot \phi .
\label{p2}
\end{equation}
Explicit expression for $j^0$ is given by,
\begin{equation}
 j^0=\frac{\beta}{2\pi}\frac{2B}{\sqrt{2cE_0}} sech( B\sqrt{\frac{\nu}{2c}}(x+2\mu kt))$$$$[\frac{2B\alpha k_{eff}}{c^2}\sqrt{\frac{\nu \mu^3}{2c\mu}} (\frac{x}{2}-\mu k t)
 tanh( B\sqrt{\frac{\nu}{2c\mu}}(x+2\mu kt)) $$$$cos \{-kx+\frac{1}{2}(-2\mu k^2 
 	+\nu \frac{B^2}{c} -2\frac{E_0}{\hbar})t+\theta \}$$$$
 - (k-\frac{2\alpha k_{eff}k^2\mu^3}{c^2})sin\{-kx+\frac{1}{2}(-2\mu k^2 +\nu \frac{B^2}{c} -2\frac{E_0}{\hbar})t+\theta \}]  .
\label{100}
\end{equation}
These fermion densities in the curved spacetime Thirring model, obtained from the optical solitons where the bosonization duality with the (Kerr non-linear) NLS  model was exploited, constitute our major result.

{\bf Discussion   :} To summarize, we have shown a duality in $1+1$-dimensions, between massive Thirring model in curved spacetime (in particular in the p resence of a black hole in semi-classical gravity) and optical solitons in Kerr non-linear models. The fermion model is mapped on to a sine-gordon model in curved spacetime via bosonization and the latter, in non-relativistic limit reproduces the optical soliton.\\
{\it Open problems}: (i) Our results, strictly speaking, are valid in the lowest order of approximation since we have used the same dictionary between fermion and boson degrees of freedom that is rigorously true in flat spacetime. In \cite{eboli} only the mapping between  coupling constants of the two models is mentioned.\\
(ii) The duality with the optical soliton model is demonstrated only in a truncated model.\\
{\it Future Directions}: The bosonization dictionary needs to be extended in the curved spacetime. The duality with the sine-Gordon model in curved spacetime, that appeared in an intermediate step requires further study.\\
 It will be interesting if some correspondence between the present result and the quantum mechanical equivalence with the Rabi model \cite{jules} can be established. The last point will be relevant in a possible experimental setup that can test the present results.\\
 One can also consider other forms of metric such as a "mock" Schwarzschild metric that is structurally same as the $3+1$-dimensional Schwarzschild metric but indeed, is not a solution of the $1+1$-dimensional Einstein equation. This form of metric has been used in curved spacetime bosonization in \cite{gass}. In this case the situation becomes more complicated   since the conformal  factor will be $x$-dependent. This will make the non-relativistic reduction tricky  and will lead to an inhomogeneous form of Kerr nonlinearity.\\
    Lastly, there exist approximate bosonization schemes in $2+1 $-dimensions and it would be very interesting to extend the present work in higher (at least in $2+1$) dimensions.


\begin{thebibliography}{99}
	
	
	\bibitem{mal} J. Maldacena et al., Phys. Rept. 323 (2000) 183.
	\bibitem{review}C. Barcelo, S. Liberati and M.  Visser, Living Rev. Relativ. 24 (2011)  3,  https://doi.org/10.12942/lrr-2011-3.
	\bibitem{jules}		J. S. Pedernales, M. Beau, S. M. Pittman, I. L. Egusquiza, L. Lamata, E. Solano, A. del Campo, Phys. Rev. Lett. 120, 160403 (2018);	(arXiv:1707.07520).
		\bibitem{col}S. R. Coleman,  Phys.Rev.D11 (1975) 2088. 
	\bibitem{mand} S. Mandelstam,Phys.Rev.D11 (1975) 3026.
		\bibitem{opsol1} 	A. Biswas, S. Konar, {\it Introduction to Non-Kerr Law Optical Solitons}, CRC Press, Boca Raton, FL, USA, 2006.
		\bibitem{zak} V. E. Zakharov and A. B. Shabat, Sov. Phys. JETP 34, 62
		(1972).
	\bibitem{seg} For a review on spatial solitons interactions, see G. I.
	Stegman and M. Segev, Science 286, 1518 (1999).
	
		\bibitem{cohen} O. Cohen, T. Carmon, M. Segev, and S. Odoulov, Opt. Lett. 27, 2031
	(2002); O. Cohen, R. Uzdin, T. Carmon, J. W. Fleischer, M. Segev, and S.
	Odoulov, Phys. Rev. Lett. 89, 133901 (2002).	
		\bibitem{bart} A. Barthelemy, S. Maneuf, and C. Froehly, Opt. Commun.
	55, 201 (1985);  J. S. Aitchison, A. M. Weiner, Y. Silberberg, M. K.
	Oliver, J. L. Jackel, D. E. Leaird, E. M. Vogel, and
	P. W. Smith, Opt. Lett. 15, 471 (1990).
	\bibitem{ait} J. S. Aitchison et al., Opt. Lett. 16, 15 (1991); M. Shalaby
	et al., Opt. Lett. 17, 778 (1992).
	\bibitem{shi} M. Shih and M. Segev, Opt. Lett. 21, 1538 (1996).
	\bibitem{tik} V. Tikhonenko, J. Christou, and B. Luther-Davies, Phys.
	Rev. Lett. 76, 2698 (1996);  H. Meng et al., Opt. Lett. 22, 448 (1997).
	\bibitem{krol} W. Krolikowski and S. A. Holmstrom, Opt. Lett. 22, 369
	(1997); 23, 97 (1998).
	\bibitem{shi1} M. Shih, M. Segev, and G. Salamo, Phys. Rev. Lett. 78,
	2551 (1997); A. Buryak et al., ibid. 82, 81 (1999)
	

	\bibitem{bj} J. E. Bjorkholm and A. Ashkin, Phys. Rev. Lett. 32,
	129 (1974); 6. V. Tikhonenko, J. Christou, and B. Luther-Davies,
	Phys. Rev. Lett. 76, 2698 (1996).
	\bibitem{du} G. Duree, J. L. Shultz, G. Salamo, M. Segev, A. Yariv,
	B. Crosignani, P. DiPorto, E. Sharp, and R. Neurgaonkar,
	Phys. Rev. Lett. 71, 633 (1993); 8. M. Shih, M. Segev, G. C. Valley, G. Salamo, B. Crosignani,
	and P. Diporto, Electron. Lett. 31, 826 (1995).
	\bibitem{ta} M. Taya, M. Bashaw, M. M. Fejer, M. Segev, and G. C.
	Valley, Phys. Rev. A 52, 3095 (1995).
	\bibitem{sw} G. A. Swartzlander, D. R. Andersen, J. J. Regan, H.
	Yin, and A. E. Kaplan, Phys. Rev. Lett. 66, 1583 (1991).
	\bibitem{pe} M. Peccianti, A. De Rossi, G. Assanto, A. De Luca, C.
	Umeton, and I. C. Khoo, Appl. Phys. Lett. 77, 7 (2000).	
	\bibitem{eboli}	O. J. P. Eboli, Phys.Rev.D, 36, (1987) 2408. 
		\bibitem{opsol}M. Inc, A. I. Aliyu,  A. Yusuf and D. Baleanu, Superlattices and Microstructures 113 (2018) 541.

	
	\bibitem{guth}M. H. Namjoo, A. H. Guth and D. I. Kaiser, Phys. Rev. D 98, 016011 (2018)
	DOI:	10.1103/PhysRevD.98.016011 .
	\bibitem{gu1} A. H. Guth, M. P. Hertzberg, and C. Prescod-Weinstein,  Phys. Rev. D 92, 103513 (2015), arXiv:1412.5930 [astro-ph.CO];  E. Braaten, A. Mohapatra, and H. Zhang,  Phys. Rev.
	D 94, 076004 (2016), arXiv:1604.00669 [hep-ph].
	\bibitem{eff} Measurable signatures of  coupling of fast and
	slow oscillating terms   in  physical systems, such as neuronal
	processes related to memory formation, are discussed in A. G. Siapas and M. A. Wilson,  Neuron 21, 1123 (1998);  Z. Clemens, M. Moll, L. Er¨oss, R. Jakus, G. R´asonyi, P. Hal´asz, and J. Born,  Eur. J. Neuroscience 33, 511 (2011).
	
	\bibitem{rb}R. Banerjee and P. Mukherjee,  	doi
	10.1016/j.nuclphysb.2018.11.002; (arXiv:1801.08373). As argued by the authors, the fact the  transformation (\ref{9}) yields the correct non-relativistic (Schroedinger) equation but reproduces the non-relativistic action modulo fast oscillating contributions might be because the complexification of $\phi$ to $\psi$ in effect doubles the degrees of freedom and introduces an additional global $U(1)$ gauge invariance (leading to conserved $\psi$ particle number). Furthermore these authors have shown that a mathematically consistent way to generate the non-relativistic Scroedinger theory from real Klein-Gordon theory is to develop the latter in light-cone framework in one dimension higher and subsequently obtain the former via a mapping of the field variables, that is similar to (\ref{9} but without complexification). In fact with this mapping the full set of conformal generators of the Schroedinger theory are recovered from  relativistic conformal generators.
	
	 \bibitem{orf}S.J. Orfanidis and R. Wang, Phys.Lett. 57B 281(1975)		\bibitem{mann} R. B. Mann, S. M. Morsink, A. E. Sikkema, and T. G.
	Steele, Phys. Rev. D 43, 3948 (1991).


\bibitem{jt}C. Teitelboim, in Quantum Theory of Gravity, edited by S.
Christensen (Hilger, Bristol, 1984), p. 327; R. Jackiw, ibid. ,
p. 403; Nucl. Phys. B252, 343 (1985).
\bibitem{su}
A. R. Brown, H. Gharibyan, H. W. Lin, L. Susskind, L. Thorlacius, and Y. Zhao
Phys. Rev. D 99, 046016 (2019).
\bibitem{syk} Holographic dual to charged SYK from 3D Gravity and Chern-Simons
A. Gaikwad, L. Kh Joshi, G. Mandal, S. R. Wadia; arXiv: 1802.07746.)
  
\bibitem{gass} R. Gass, Phys. Rev. D 27, 2893 (1983).
\end{thebibliography}
\end{document}